\documentclass[11pt]{article}
\usepackage[a4paper, total={6in, 8in}]{geometry}
\usepackage{amsmath}
\usepackage{amssymb}
\usepackage{hyperref}
\usepackage{xfrac}
\usepackage[]{matlab-prettifier}
\usepackage{float}
\usepackage{algpseudocode}
\usepackage{array}
\usepackage{pdfpages}
\usepackage{listings}
\usepackage{bussproofs}

\renewcommand\thesection{\arabic{section}.}

\title{A Coq implementation of a Theory of Tagged Objects}
\author{Matthew Gates and Alex Potanin}
\date{}
\begin{document}
	\maketitle
	
\begin{abstract}
We present a first step towards the Coq implementation of the Theory of Tagged Objects formalism. The concept of tagged types is encoded, and the soundness proofs are discussed with some future work suggestions.
\end{abstract}

    \section{INTRODUCTION}
    
    Modelling a statically typed object-oriented class-based language with a dynamic class hierarchy in foundational type theory has several benefits. In particular, it is more flexible as it allows reasoning about languages with dynamic class hierarchies and other more dynamic compositions. The type theory approach allows better reasoning about classes and comparing classes to ideas in different programming paradigms.
    
    We begin with the core language presented in A Theory of Tagged Objects (ToTO) \cite{lee_et_al:LIPIcs.ECOOP.2015.174}. We choose not to implement the source class-based language or the translations between the two languages. In ToTO a class from the source language is modelled as a sum of the classes tag and a constructor function to create the objects.

    We first present a slightly modified type system from the one presented in the ToTO \cite{lee_et_al:LIPIcs.ECOOP.2015.174}, The main contribution is encoding this type system into the Coq Proof assistance and declaration of the type soundness theorems. The encoded language is able to express terms from the ToTO core language.
    
    \section{ToTO TYPE SYSTEM}
    
     \begin{figure}
        \begin{lstlisting}
    $n$ $::=$ $x$ $|$ $c$ $|$ Fst$(n)$ $|$ Unfold$(n)$
    $e$ $::=$ Newtag$[\tau]$ $|$ Subtag$[\tau]$ $|$ New$(n)\{e\}$ $|$ match$\{e\}(n)(x)\{e\}\{e\}$ $|$ /$x:\tau,e$ 
        $|$ $e$ $e$ $|$ $\{\overline{f=e}\}$ $|$ $e$ proj $f$ $|$ Let $x$ be $e$ in $e$ $|$ Fix$\{e\}$
        $|$ Fold$[t.\tau]\{e\}$ $|$ Unfold$\{e\}$ $|$ $\langle e,e\rangle$ $|$ Fst$\{e\}$ $|$ Snd$\{e\}$ $|$ < > $|$ $n$
    $\tau$ $::=$ $Tag(\tau)$ $|$ Tagged$(n)$ $|$ Prod$[x:\tau],\tau$ $|$ $\{\overline{f:\tau}\}$ $|$ mu$(t):\tau$ $|$ Sum$[x:\tau]\tau$ $|$ Top $|$ $t$
    $Tag(\tau)$ $::=$ Tag$[\tau]$  $|$ Tag$[\tau]$extends$(n)$
    $\Gamma$ $::=$ $\epsilon$ $|$ $\Gamma,x:\tau$
    $\Sigma$ $::=$ $\epsilon$ $|$ $\Sigma,c \sim\tau$
    $\Delta$ $::=$ $\epsilon$ $|$ $\Delta,t<:t$
        \end{lstlisting}
        \caption{Theory of Tagged objects type encoding}
        \label{fig:Source}
    \end{figure}
    
    The syntax is given in Figure \ref{fig:Source}. Other than the notations the only difference to the core language of ToTO presented in \cite{lee_et_al:LIPIcs.ECOOP.2015.174} is the \verb|Fix| expression as opposed to the recursive let binding. This is because of the choice to define the recursive let binding as a derived form of the let binding and \verb|Fix|$\{e\}$.
    
    Beginning from the start of the syntax we have the names $n$ first names are variables $x$, tags $c$ or unfolded out of other names as in \verb|Fst|$(n)$ or \verb|Unfold|$(n)$.
    
    Up next are types $\tau$ are the Tag types, tagged types, dependent product and sums, records, iso recursive types, top types and type variables. The dependent types are necessary in the ToTO language as they allow the translation to and from a class based core language. The tag types $Tag(\tau)$ can either just be a defined by a type as in  \verb|Tag|$[\tau]$ or as a tag type extended from a name as in \verb|Tag|$[\tau]$\verb|extends|$(n)$.
    
    The terms $e$ cover tag creation and matching, abstraction, application, records, let bindings, fix statements, folding and unfolding recursive statements, pair creation and separation, units and, names.
    
    In addition to a typing context $\Gamma$ and a subtyping context $\Delta$. The $\Sigma$ contexts acts more as store linking tags and types. That is if a given type $\tau$ is tagged with $c$ then under $\Sigma$, $c$ will map to $\tau$.
    
    \begin{figure}
        \begin{prooftree}
            \AxiomC{}
            \RightLabel{ST-Reflexive}
            \UnaryInfC{$\Delta|\Gamma\vdash_\Sigma t<:t$}
        \end{prooftree}
        \begin{prooftree}
            \AxiomC{$\Delta|\Gamma\vdash_\Sigma \tau_1<:\tau_2$}
            \AxiomC{$\Delta|\Gamma\vdash_\Sigma \tau_2<:\tau_3$}
            \RightLabel{ST-Transitive}
            \BinaryInfC{$\Delta|\Gamma\vdash_\Sigma \tau_1<:\tau_3$}
        \end{prooftree}
        \begin{prooftree}
            \AxiomC{$t<:t'\in\Delta$}
            \RightLabel{ST-Amber-1}
            \UnaryInfC{$\Delta|\Gamma\vdash_\Sigma t<:t'$}
        \end{prooftree}
        \begin{prooftree}
            \AxiomC{$\Delta,t<:t'|\Gamma\vdash_\Sigma\tau<:\tau'$}
            \RightLabel{ST-Amber-2}
            \UnaryInfC{$\Delta|\Gamma\vdash_\Sigma$ mu$( t):\tau<:$mu$(t'):\tau'$}
        \end{prooftree}
        \begin{prooftree}
            \AxiomC{}
            \RightLabel{ST-Record-1}
            \UnaryInfC{$\Delta|\Gamma\vdash_\Sigma \{f_i:\tau_i^{i\in 1...n+k}\}<: \{f_i:\tau_i^{i\in 1...n}\}$}
        \end{prooftree}
        \begin{prooftree}
            \AxiomC{for each $i$  $\Delta|\Gamma\vdash_\Sigma \tau_i<:\tau_i'$}
            \RightLabel{ST-Record-2}
            \UnaryInfC{$\Delta|\Gamma\vdash_\Sigma \{f_i:{\tau_i}^{i\in 1...n}\}<: \{f_i:{\tau_i'}^{i\in 1...n}\}$}
        \end{prooftree}
        \begin{prooftree}
            \AxiomC{$\{l_j:{\rho_j}^{j\in 1...n}\}$ is a permutation of $\{f_i:{\tau_i}^{i\in 1...n}\}$}
            \RightLabel{ST-Record-3}
            \UnaryInfC{$\Delta|\Gamma\vdash_\Sigma \{l_j:{\rho_j}^{j\in 1...n}\}<: \{f_i:{\tau_i'}^{i\in 1...n}\}$}
        \end{prooftree}
        \begin{prooftree}
            \AxiomC{$\Delta|\Gamma\vdash_\Sigma \tau_3<:\tau_1$}
            \AxiomC{$\Delta|\Gamma,x:\tau_3\vdash_\Sigma \tau_2<:\tau_4$}
            \RightLabel{ST-App}
            \BinaryInfC{$\Delta|\Gamma\vdash_\Sigma $Prod$[x:\tau_1],\tau_2<:$Prod$[x:\tau_3],\tau_4$}
        \end{prooftree}
        \begin{prooftree}
            \AxiomC{$\Gamma\vdash_\Sigma n:$Tag$[\tau]$extends$(n')$}
            \RightLabel{ST-Tag-1}
            \UnaryInfC{$\Delta|\Gamma\vdash_\Sigma$ Tagged$(n)<:$ Tagged$(n')$}
        \end{prooftree}
        \begin{prooftree}
            \AxiomC{$\Delta|\Gamma\vdash_\Sigma$ Tagged$(n)<:$ Tagged$(n')$}
            \RightLabel{ST-Tag-2}
            \UnaryInfC{$\Delta|\Gamma\vdash_\Sigma$ Tag$[\tau]$extends$(n)<:$ Tag$[\tau]$extends$(n')$}
        \end{prooftree}
        \begin{prooftree}
            \AxiomC{}
            \RightLabel{ST-Tag-3}
            \UnaryInfC{$\Delta|\Gamma\vdash_\Sigma\tau$ Tag$[\tau]$extends$(n)<:$ Tag$[\tau]$}
        \end{prooftree}
        \caption{Subtyping Rules}
        \label{fig:Subtype}
    \end{figure}
    
    \subsection{Subtyping}
    
    We define the Subtyping rules of the language in Figure \ref{fig:Subtype}. The subtyping rules make judgement in relation to all three contexts in particular the $\Delta$ context links type variables and is used in the \textit{ST-Amber-1} and \textit{ST-Amber-2} rules. The Amber rules are used for making judgements on the iso-recursive types and are based on\cite{DBLP:conf/litp/Cardelli85}.
    
    The sybtyping of records is the regular subtyping for records. The \textit{ST-App} rule is only a minor difference from the regular judgement. Because the ToTO type system has a dependent product type we must use contravariant rule Which utilises the $\Gamma$ context to associate the variable $x$ with the type $\tau_3$.
    
    The subtyping of tagged types nominal and is handled by the rule \textit{ST-Tag-1}. This works as expected if $n$ has the type of a subtag of $n'$ then a tagged type with tag $n$ is a subtype of a tagged type with tag $n'$. The general reflexivity and transitivity rules handle subtagging reflexivity and transitivity.
    
    The subtyping of tag types is a little more interesting. The first rule \textit{ST-Tag-2}. Handles the case of when is a tag type a subtype of tag type with supertag $n'$. That is subtypes of $\tau$\verb| tag extends |$n'$. A tag type $\tau$\verb| tag extends |$n$ is a subtype of  $\tau$\verb| tag extends |$n'$ when $n$ is a subtag of $n'$, this is \textit{ST-Tag-2}. For the final judgement we want a tag with no supertag to be a supertype of any tag with a supertag that is as in \textit{ST-Tag-3}.
    
    \subsection{Typing}
    
    The typing rules related to the creation and interaction of tags and those related to the dependent and recursive types are listed in Figure \ref{fig:Type}. Type judgements have the form $\Gamma \vdash_\Sigma e:\tau$.
    
    \begin{figure}
        \begin{prooftree}
            \AxiomC{$\Gamma \vdash_\Sigma e_1:$ Prod$[x:\tau],\tau'$}
            \AxiomC{$\Gamma \vdash_\Sigma e_2:\tau$}
            \RightLabel{App}
            \BinaryInfC{$\Gamma\vdash_\Sigma e1e2:[e2/x]\tau'$}
        \end{prooftree}
        \begin{prooftree}
            \AxiomC{$\Gamma \vdash_\Sigma e:\tau$}
            \AxiomC{$\epsilon|\Gamma \vdash_\Sigma \tau<:\tau'$}
            \RightLabel{Sub}
            \BinaryInfC{$\Gamma\vdash_\Sigma e:\tau'$}
        \end{prooftree}
        \begin{prooftree}
            \AxiomC{$\Sigma(c)=\tau$}
            \RightLabel{CVar}
            \UnaryInfC{$\Gamma\vdash_\Sigma c:\tau$}
        \end{prooftree}
        \begin{prooftree}
            \AxiomC{}
            \RightLabel{Cls-I}
            \UnaryInfC{$\Gamma\vdash_\Sigma $ Newtag$[\tau]:$Tag$[T]$}
        \end{prooftree}
        \begin{prooftree}
            \AxiomC{$\Gamma \vdash_\Sigma n:Tag(\tau)$}
            \AxiomC{$\epsilon|\Gamma \vdash_\Sigma \tau'<:\tau$}
            \RightLabel{CCls-I}
            \BinaryInfC{$\Gamma\vdash_\Sigma$ Subtag$[\tau'](n):$ Tag$[\tau']$extends$(n)$}
        \end{prooftree}
        \begin{prooftree}
            \AxiomC{$\Gamma \vdash_\Sigma n:Tag(\tau)$}
            \AxiomC{$\Gamma \vdash_\Sigma e:\tau$}
            \RightLabel{Tag-I}
            \BinaryInfC{$\Gamma\vdash_\Sigma$ New$(n)\{e\}:$ Tagged$(n)$}
        \end{prooftree}
        \begin{prooftree}
            \AxiomC{$\Gamma \vdash_\Sigma e1:$ Tagged$(n')$}
            \noLine
            \UnaryInfC{$\Gamma,x:$ Tagged$(n) \vdash_\Sigma e_2:\tau$}
            \AxiomC{$\Gamma \vdash_\Sigma$ Tagged$(n')\updownarrow$ Tagged$(n)$}
            \noLine
            \UnaryInfC{$\Gamma \vdash_\Sigma e3:\tau$}
            \RightLabel{Match}
            \BinaryInfC{$\Gamma\vdash_\Sigma$ Match$\{e_1\}(n)(x)\{e_2\}\{e_3\}:\tau$}
        \end{prooftree}
        \begin{prooftree}
            \AxiomC{$\Gamma \vdash_\Sigma e1:\tau_1$}
            \AxiomC{$\Gamma \vdash_\Sigma e2:[e1/x]\tau_2$}
            \RightLabel{$\Sigma$-I}
            \BinaryInfC{$\Gamma\vdash_\Sigma \langle e1,e2\rangle:\Sigma_{x:\tau_1}\tau_2$}
        \end{prooftree}
        \begin{prooftree}
            \AxiomC{$\Gamma \vdash_\Sigma e:$ Sum$[x:\tau_1]\tau_2$}
            \RightLabel{$\Sigma$-E$_1$}
            \UnaryInfC{$\Gamma\vdash_\Sigma $ Fst$\{e\}:\tau_1$}
        \end{prooftree}
        \begin{prooftree}
            \AxiomC{$\Gamma \vdash_\Sigma e:$ Sum$[x:\tau_1]\tau_2$}
            \RightLabel{$\Sigma$-E$_2$}
            \UnaryInfC{$\Gamma\vdash_\Sigma $ snd$\{e\}:[$Fst$\{e\}/x]\tau_2$}
        \end{prooftree}
        \caption{Type Judgements}
        \label{fig:Type}
    \end{figure}
    
    In this the \textit{App} and \textit{$\Sigma$} rules all rely on a substitution into types. However in the syntax a type cannot depend on a general expression and can only depend on a name. Therefore the type substitution is only valid when the expression $e$ reduces to a name which has no effects $x$ is not free in the type $\tau'$ being substituted into. In all other cases the substitution is not defined and the \textit{App} and \textit{$\Sigma$} rules cannot be applied.
    
    The remaining rules include subsumption, a rule to look up tag values. The rules include very basic facts such as the creation of a new tag has the type of a tag and the creation a new subtag has the type of a new subtag. Similarly the creation of a new tagged expression has the type of a tagged type.
    
    The \textit{Match} rule is slightly more involved with a match expression having the type $\tau$ when both $e_2$ and $e_3$ have type $\tau$ and $e_1$ is tagged with a name $n$ with a mutual supertag as the variable $x$ has. The $\updownarrow$ denotes two types having a shared supertype. That is there exists some type that both types are subtypes of, formally.
    
    \[\Gamma\vdash_\Sigma\tau_1\updownarrow\tau_2=\exists\tau_p.\epsilon|\Gamma\vdash_\Sigma\tau_1<:\tau_p\land\epsilon|\Gamma\vdash_\Sigma\tau_2<:\tau_p\]
    
    As mentioned before the sum type judgements are slightly different because they must factor in the possible dependency of $\tau_2$ on $x$. Similar to for \textit{App} this means including type substitution into the second type when making type judgements of the pair in $\Sigma$-I. For making judgements about the \verb|Snd| projection in $\Sigma$-E$_2$ we must utilise the substitution into types.
    
    \subsection{Dynamics} 
    
    There are two parts to the dynamics of the ToTO language the value judgements and the small step evaluation rules. Both parts rely on a hierarchical store to track the generated tag values. We construct the hierarchical store $S$ as a set of paths beginning at tag values $c$ and pointing to a possibly empty paths $p$ the definitions are listed in figure \ref{fig:store} 
    
    \begin{figure}
        \begin{align*}
            S &::=\epsilon|S,c\rightsquigarrow p\\
            p &::=\epsilon|c\rightsquigarrow p
        \end{align*}
        \begin{prooftree}
            \AxiomC{}
            \UnaryInfC{$c\in c\rightsquigarrow p$}
        \end{prooftree}
        \begin{prooftree}
            \AxiomC{$c\in p$}
            \AxiomC{$c\neq c'$}
            \BinaryInfC{$c\in c'\rightsquigarrow p$}
        \end{prooftree}
        \caption{Store Definitions}
        \label{fig:store}
    \end{figure}
    
    The dynamics of the language are judgements based on the hierarchical store. with value judgements having the form $S$ $|$ $e$\verb| val| which is stating that $e$ is a value under the context of the hierarchical store $S$. The small step reductions are made based on the Hierarchical store $S$ and step from an expression $e$ to another expression $e'$ while updating $S$ to $S'$, denoted by $S$ $|$ $e\mapsto$ $S'$ $|$ $e'$.
    
    The one value judgement that we need to examine is the \textit{C-V} rule below
    
    \begin{prooftree}
        \AxiomC{$c\in S$}
        \RightLabel{C-V}
        \UnaryInfC{$S$ $|$ $c$ val}
    \end{prooftree}
    
    This is the judgement to say that if we have a given tag value $c$ in a store then that tag value is a value.
    
    \begin{figure}
        \begin{prooftree}
            \AxiomC{$c\notin S$}
            \RightLabel{$\mapsto$ CLS}
            \UnaryInfC{$S$ $|$ Newtag$[\tau]\mapsto S,c$ $|$ $c$}
        \end{prooftree}
        \begin{prooftree}
            \AxiomC{$c'\notin S$}
            \RightLabel{$\mapsto$ CCLS}
            \UnaryInfC{$S,c\rightsquigarrow p$ $|$ Newtag$[\tau]\mapsto S,c'\rightsquigarrow(c\rightsquigarrow p),c\rightsquigarrow p$ $|$ $c'$}
        \end{prooftree}
        \begin{prooftree}
            \AxiomC{$S$ $|$ $e\mapsto S'$ $|$ $e'$}
            \RightLabel{$\mapsto$ New}
            \UnaryInfC{$S$ $|$ New$(n)\{e\}\mapsto S'$ $|$ New$(n)\{e'\}$}
        \end{prooftree}
        \begin{prooftree}
            \AxiomC{$S$ $|$ $e\mapsto S'$ $|$ $e'$}
            \RightLabel{$\mapsto$ Match}
            \UnaryInfC{$S$ $|$ Match$\{e\}(n)(y)\{e_2\}\{e_3\}\mapsto S'$ $|$ Match$\{e'\}(n)(y)\{e_2\}\{e_3\}$}
        \end{prooftree}
        \begin{prooftree}
            \AxiomC{$S,c\rightsquigarrow p$ $|$ $e$ val}
            \AxiomC{$c'\in c\rightsquigarrow p$}
            \RightLabel{$\mapsto$ MatchSuc}
            \BinaryInfC{$S,c\rightsquigarrow p$ $|$ Match$\{$New$(c)\{e\}\}(c')(y)\{e_2\}\{e_3\}\mapsto S',c\rightsquigarrow p$ $|$ $[y:=$New$(c)\{e\}]e_2$}
        \end{prooftree}
        \begin{prooftree}
            \AxiomC{$S,c\rightsquigarrow p$ $|$ $e$ val}
            \AxiomC{$c'\notin c\rightsquigarrow p$}
            \RightLabel{$\mapsto$ MatchFail}
            \BinaryInfC{$S,c\rightsquigarrow p$ $|$ Match$\{$New$(c)\{e\}\}(c')(y)\{e_2\}\{e_3\}\mapsto S',c\rightsquigarrow p$ $|$ $e_3$}
        \end{prooftree}
        \begin{prooftree}
            \AxiomC{$S$ $|$ $e\mapsto S'$ $|$ $e'$}
            \RightLabel{$\mapsto$ Untag1}
            \UnaryInfC{$S$ $|$ Extract$\{e\}\mapsto S'$ $|$ Extract$\{e'\}$}
        \end{prooftree}
        \begin{prooftree}
            \AxiomC{$S$ $|$ $e$ val}
            \RightLabel{$\mapsto$ Untag2}
            \UnaryInfC{$S$ $|$ Extract$\{$New$($\_$)\{e\}\}\mapsto S'$ $|$ $e$}
        \end{prooftree}
        \caption{Reduction rules}
        \label{fig:red}
    \end{figure}
    
    Moving on we examine the small step reductions implemented in Figure \ref{fig:red} these are the small step reductions for the tagged types and are taken from from the theory of tagged objects. These are just the judgements relating to the reduction of the tag expressions.\\ First we are particularly interested in the $\mapsto$ CLS and $\mapsto$ CCLS rules. In these rules we check if a given tag value already exists in the hierarchy. In the case it does not the step can be made to the new tag value while adding the tag value to the store.
    
    The next rules to discuss after this one are the $\mapsto$ MatchSuc and $\mapsto$ MatchFail. They work by checking if a given tag value $c'$ is in the path of $c$, which is equivalent to saying that $c'$ is a supertag of $c$. Then if $c'$ does appear as a supertag. Then it would be expected if the \verb|Match| were to work properly that $e_2$ would be the result and indeed under $\mapsto$ MatchSuc that is the result. $\mapsto$ MatchFail reduces to $e_3$ under the case where $c'$ is not a supertag of $c$.
    
    We've now covered all the new syntax and semantics of the language.
    
    \section{ToTO COQ ENCONDING}
    
    The encoding of the theory of tagged objects has two parts. In the first part is a set of definitions for the syntax, reduction and typing rules. Following that is an encoding of the theorems of type soundness. These encodings do not include a proof and are simply stated.
    
    \subsection{Definitions}
    
    The definitions of the language is split into several parts. The first part corresponds to the syntax of the language relating to names, terms, types, stores and contexts of the language. This is followed by the subtyping relation and implementation of substitutions. The back half explores the type relation and then the reduction rules.
    
    \subsubsection{Syntax Definition}
    
    Before we begin defining the specific syntax of constructions in the language we discuss the general syntax in the language we use several  wrappers. \verb|"<{ e }>"| to denote an expression that make up a program. We use \verb|"<{{ T }}>"| to denote a type being declared.
    
    We define most of the syntax of the language inductively beginning with the names. We can construct names in any one of 4 ways. The \verb|n_var| construction is to implement using a variable as name. We represent variables in the implementation as a \verb|string| we later implement it as a coercion so that w. The \verb|n_tag| construction is a general construction for an existing tag value in notations we define the keyword \verb|"'c'"| to represent the construction.
    
    \begin{verbatim}
    Inductive name : Type :=
        | n_var     : string -> name
        | n_tag     : name
        | n_fst     : name -> name
        | n_unfold  : name -> name.
    \end{verbatim}
    The last two constructions are for projecting and unfolding existing names and we define the below notations as syntax for using these constructions in the programming language.
    \begin{verbatim}
    Notation "Fst( n )" := (n_fst n) (in custom toto_ty at level 4).
    Notation "Unfold( n )" := (n_unfold n) (in custom toto_ty at level 4).
    \end{verbatim}
    We will next the discuss the type encoding in 5 sections beginning with the way that tag and tagged types are encoded
    \begin{figure}[H]
    \begin{verbatim}
    Inductive ty : Type :=
        | ty_tag1     : ty -> ty
        | ty_tag2     : ty -> name -> ty
        | ty_tagged   : name -> ty
    \end{verbatim}
    \end{figure}
    
    The first two are the tag types $Tag(\tau)$ in the ToTO language are defined along with the notations \verb|"Tag[T]"| for $\tau$ \verb|tag| and \verb|"Tag[T]Extends(n)"| for $\tau$ \verb|tag extends| $n$ these two are to denote the type of tags being used to tag types \verb|T| and tag a type \verb|T| as an extension of the tag \verb|n|. The third construction is the type of a type that has been tagged with the name \verb|n| and is defined along with the notation \verb|Tagged(n)| for \verb|tagged| $n$.
    
    We next discuss the two dependent types. These are the dependent sum and dependent product. The syntax is defined below again using strings to implement variables.
    
    \begin{verbatim}
        | ty_prod     : string -> ty -> ty -> ty
        | ty_sum      : string -> ty -> ty -> ty
    \end{verbatim}
    
    These types are also defined along with with notations. The dependent product is denoted by \verb|"Prod[ x : T1 ], T2"|. This notation mimics the Capital pi $(\Pi)$ syntax replacing the capital pi with the keyword \verb|Prod|. The sum type syntax is much the same deriving from the $\Sigma$ notation but using the \verb|Sum| keyword.
    
    The record types are defined in the same way as in Software Foundations, Programming languages fundamentals\cite{Pierce:SF2}. The constructors for this are below, one for the \verb|"nil"| empty record, and another for a recursive call with notation \verb|"i;T;;R"| This construction necessitates a check for being well formed since we allow \verb|R| to be any type when we only actually allow it to be a record type these two checks are the propositional logic  are \verb|record_ty| and \verb|wellformed_ty|.
    
    It is important to note that this definition of a Record is recursive and so to check if a given variable is in the record we follow Programming Languages Foundations in defining a recursive function \verb|tyLookup|.
    
    Moving to the last grouping is the recursive, top and unit types. Their implementation is below.
    
    \begin{verbatim}
        | ty_isorec   : ty -> ty -> ty
        | ty_top      : ty
        | ty_var     : string ->  ty.
    \end{verbatim}
    
    The recursive type has the notation \verb|mu(t):T|.
    
    The unit type is implemented as a generic string as it allows the use of a string as a type in particular in relation to the contexts that we will we define later.
    
    The next part of the syntax is the terms of the language that allow programs to be written, We again split into several parts to discuss this first beginning with the terms relating to tagging. There are 5 term constructions relating to tagging listed below.
    
    \begin{verbatim}
        Inductive tm : Type :=
            | tm_newtag   : ty -> tm
            | tm_subtag   : ty -> name -> tm
            | tm_new      : name -> tm -> tm
            | tm_match    : tm -> name -> string -> tm -> tm -> tm
            | tm_extract  : tm -> tm
    \end{verbatim}
    
    These constructions directly line up with the constructions in the language relating to the creation and interaction of tags all constructions line up directly with terms from the Theory of Tagged objects. The first three relating to the creation of tagged expressions are defined along with the notations \verb|"NewTag[T]"|, \verb|"SubTag[T](n)"| and \verb|"New{e}(n)"|.
    The last two for the interactions are defined with the notations \verb|"Match{ e1 }( n )( x ){ e2 }{ e3 }"| and \verb|"Extract{ e }"|.
    
    The next two terms are pulled directly from simply typed lambda calculus abstraction and application. The notation for application is the same as it is for standard simply typed lambda calculus. Abstraction is defined using the notation \verb|"/x:T,e"|.
    
    Moving on we define the terms relating to records.
     
    \begin{verbatim}
        | tm_rnil     : tm
        | tm_rcons    : string -> tm -> tm -> tm
        | tm_proj     : tm -> string -> tm
    \end{verbatim}
    
    The notations for these are \verb|"nil"|, \verb|"i;=e1;;e2"| and \verb|"e proj f"|. Clearly the \verb|tm_rcons| construction suffers from the same issues as its type counterpart where \verb|e2| can be a non record term. To combat this we \verb|record_tm| to check if a given term is a record term, that is either \verb|tm_rnil| or \verb|tm_rcons|. This follows Programming languages foundations\cite{Pierce:SF2}.
    
    The next two constructions are the let bindings and fix construction. We do this in two stages first defining a let term \verb|tm_let : string -> tm -> tm -> tm| along with the notation \verb|"Let x be e1 in e2"|. Following that we define the fix expression and then following Pierce we define the recursive let binding as a derived form, This does mean we have a slightly different definition where we state the type of \verb|x| in the term for the recursive let binding than in the ToTO \cite{lee_et_al:LIPIcs.ECOOP.2015.174} core language. Despite this the same programs should be able to be implemented. Therefore we define fix as \verb|tm_fix : tm -> tm| with notation \verb|"Fix{e}"|. Then define the recursive let bindings as below.
    \begin{verbatim}
        Definition tm_letrec (x:string) (T:ty) (e1:tm) (e2:tm) : tm :=
            <{Let x be Fix{/x:T,e1} in e2}>.
    \end{verbatim}
    We then state the notation
    \begin{verbatim}
        "LetRec x:T be e1 in e2" := tm_letrec x T e1 e2
    \end{verbatim}
    
    The next 3 terms relate to the sum type and are as expected with notation \verb|"Fst{e}"| and \verb|"Snd{e}"| being the notation for projecting first and second terms. The \verb|tm_prod : tm -> tm -> tm| construction for creating such pairs also has the notation \verb|"<e1,e2>"| recalling that this is the dependent sum type. After this we just have the unit term along with the notation \verb|"< >"|. The name term is simply a coercion so any valid construction of a name is also a valid term.
    
    The next important construction is the implementation of contexts and stores. The contexts and the hierarchical stores are defined differently so we will begin with the contexts. All three contexts are defined the same. That is as a partial function from strings to types as below along with a definition of an empty context and a context updating function
    
    \begin{verbatim}
        Definition context := string -> (option ty).
        
        Definition c_empty : context :=
            (fun _ => None).
        
        Definition c_update (G:context) (x:string) (T : option ty) :=
            fun x' => if String.eqb x x' then  T else G x'.
    \end{verbatim}
    
    The context implements follows on from the implementation of the simply typed lambda calculus in Programming languages Foundations \cite{Pierce:SF2}. For the other two we are again linking some form of variable to a type and so the same implementation makes sense. Because the ToTO language has different notations for the three different contexts it made sense to not define a specific notation for the contexts since we would be defining three separate notations for the same thing as far as the Coq handler was concerned.
    
    The store type is to record the tags generated by the program dynamically since a function may generate a tag and then be called an arbitrary number of times. We want the hierarchical store to be a set of paths each starting from a tag value $c$ and pointing to a path $p$.
    
    The store type implementation is slightly more involved. I chose to define the store type recursively in two parts with a path being a list of names culminating in an empty path. While the store is a list of such paths. 
    
    \begin{verbatim}
        Inductive path : Type :=
            | path_path : name -> path -> path
            | path_empty : path.
        
        Inductive store : Type :=
            | store_empty : store
            | store_cons : path -> store -> store.
    \end{verbatim}
    
    For the full implementation, we must define functions to determine whether a given name is in the store type to do this we define the predicate relations \verb|nameinstore| and \verb|nameninstore| These rely on similarly defined predicates for determining if a name is a path with notations \verb|"x inp p"| and \verb|"x ninp p"|. The definitions for stores are below.
    
    \begin{verbatim}
        Inductive nameinstore : name -> store -> Prop :=
            | s_id : forall x S p,
                x inp p ->
                nameinstore x (store_cons p S)
            | s_step : forall x p S,
                x ninp p ->
                nameinstore x S ->
                nameinstore x (store_cons p S).
        
        Notation "x 'ins' S" := (nameinstore x S) (at level 101,
        S custom toto_path at level 0, right associativity).
        
        Inductive namenotinstore : name -> store -> Prop :=
            | ns_id : forall x,
                namenotinstore x store_empty
            | ns_step : forall x p S,
                x ninp p ->
                namenotinstore x S ->
                namenotinstore x (store_cons p S).
        
        Notation "x 'nin' S" := (namenotinstore x S) (at level 101,
        S custom toto_path at level 0, right associativity).
    \end{verbatim}
    
    The stores are necessary for the evaluation and value predicates defined later. In addition we do define notation for the stores and paths which can be seen later which ease readability of the evaluation rules later. For now we continue by defining the subtyping relation.
    
    \subsubsection{Subtyping}
    
    The type of the subtyping relation is given below. There are many cases to consider for when two types could be subtypes of others but. There are eight cases in particular which have some more details required.
    
    \begin{verbatim}
        Inductive subtype : context -> context -> context -> ty -> ty -> Prop :=
    \end{verbatim}
    This is defined using the notations below
    \begin{verbatim}
        "D ! G |- S , t <: t'" := subtype D G S t t'
    \end{verbatim}
    
    This expresses that under the contexts subtyping context \verb|D|, the typing context \verb|G| and the tag linking context \verb|S| then the type \verb|t| is a subtype of the type \verb|t'|.
     
    \begin{prooftree}
        \AxiomC{H1}
        \AxiomC{H2}
        \RightLabel{Rule}
        \BinaryInfC{C}
    \end{prooftree}
    
    In general, throughout the bulk of rules for both subtyping, typing and reductions, we implement the above rule as below.
    
    \begin{verbatim}
        | s_rule : forall H1 H2 C,
            H1 ->
            H2 ->
            C
    \end{verbatim}
    
    When working with types in our implementation, we cannot ensure that a general type is actually a well-formed type in the ToTO language. Therefore, for each predicate, we must check the types are wellformed using the \verb|wellformed_ty| function.
    
    As we move on, we first examine the cases \verb|s_amb1| and \verb|s_amb2| these are the amber rules.
    
    The next cases of interest are the record cases \verb|s_rcd1|, \verb|s_rcd2| and \verb|s_rcd3|. These are of interest as in contrast to all other rules these are not direct implementations of the rules from ToTO type system. This difference is due to the way records are defined recursively, these rules match the Programming Languages Foundation implementation\cite{Pierce:SF2}. \verb|s_rcd1| is making the statement that any record is subtype of the empty record. that is implemented.
    
    \begin{verbatim}
        | s_rcd1 : forall D G S i T R,
            wellformed_ty <{{ i ; T ;; R }}> ->
            D ! G |- S , <{{ i ; T ;; R }}> <: <{{ nil }}>
    \end{verbatim}
    
    This rule along with the rule \verb|s_rcd2| together implement \textit{ST-Record-1} and \textit{ST-Record-2} from the ToTO type system \verb|s_rcd2| is implemented as below.
    
    \begin{verbatim}
        | s_rcd2 : forall D G S i1 i2 T1 T2 R1 R2,
            wellformed_ty <{{ i1 ; T1 ;; R1 }}> ->
            wellformed_ty <{{ i2 ; T2 ;; R2 }}> ->
            D ! G |- S , T1 <: T2 ->
            D ! G |- S , R1 <: R2 ->
            D ! G |- S , <{{ i1 ; T1 ;; R2 }}> <: <{{ i2 ; T2 ;; R2 }}>
    \end{verbatim}
    
    These together implement as \verb|s_rcd2| can be applied with \verb|T1=T2| until \verb|R2=<{{nil}}>| when \verb|s_rcd1| can be applied. The final record rule is \verb|s_rcd3|
    
    \begin{verbatim}
     | s_rcd3 : forall D G S i1 i2 T1 T2 R,
         wellformed_ty <{{ i1 ; T1 ;; i2 ; T2 ;; R }}> ->
         i1 <> i2 ->
         D ! G |- S , <{{ i1 ; T1 ;; i2 ; T2 ;; R }}> <: <{{ i2 ; T2 ;; i1 ; T1 ;; R }}>
    \end{verbatim}
    
    This rule finishes off the implementation of the ToTO type system record rules.
    
    The last unique part of the subtyping relation implementation of interest is the tagging subtyping implementation. This is listed below.
    
    \begin{verbatim}
        | s_tag2 : forall D G S n n' T,
            wellformed_ty T ->
            D ! G |- S , Tagged( n ) <: Tagged( n' ) ->
            D ! G |- S , Tag[ T ]Extends( n ) <: Tag[ T ]Extends( n' )
        | s_tag3 : forall D G S n T,
            wellformed_ty T ->
            D ! G |- S , Tag[ T ]Extends( n ) <: Tag[ T ]
    \end{verbatim}
    
    There is nothing of particular interest in these two rules. The interest is in the implementation of \textit{ST-Tag-1} which states that if \verb|n| has type \verb|Tag[ T ]Extends( n' )| then \verb|Tagged( n )| is a subtype of \verb|Tagged( n' )|. The issue is that we do not have a defined typing relation yet. A further issue to this is that the typing relation depends on having a defined subtyping relation. To counteract this we seperately define the rule below.
    
    \begin{verbatim}
        Definition s_tag1 : Prop := forall D G S (n:name) n' T,
            wellformed_ty T ->
            G |-- S, n :: <{{ Tag[ T ]Extends( n' )}}> ->
            D ! G |- S , Tagged( n ) <: Tagged( n' ).
    \end{verbatim}
    
    \subsubsection{Expression Substitution}
    
    Following from Programming languages Foundations we define the term substitution with a function. This function has the type below.
    
    \begin{verbatim}
        Fixpoint subst (x:string) (s:tm) (t:tm) : tm :=
    \end{verbatim}
    
    The general substitution is modelled off of Programming languages foundations with just a couple of cases different and a few new cases. Beginning with the tagging cases these are listed below and the only case of interest in these is the \verb|Match| case since all others have no dependence on a variable.
    
    \begin{verbatim}
        | <{NewTag[T]}> =>
            <{NewTag[T]}> 
        | <{SubTag[T](n)}> =>
            <{SubTag[T](n)}>
        | <{New{e}(n)}> =>
            <{New{[x:=s] e }(n)}>
        | <{Match{e1}(n)(y){e2}{e3} }> =>
          if String.eqb x y then 
            <{Match{e1}(n)(y){e2}{[x:=s]e3} }>
          else 
            <{Match{[x:=s]e1}(n)(y){[x:=s]e2}{[x:=s]e3} }>
        | <{Extract{e} }> =>
            <{Extract{[x:=s]e} }>
    \end{verbatim}
    
    The \verb|Match| case seems too simple.
    The only other case of interest is the \verb|<{n}>| case which is implemented below.
    
    \begin{verbatim}
        | <{n}> =>
            match n with
            | n_var y =>
                if String.eqb x y then s else t
            | _ => <{n}>
            end
    \end{verbatim}
    
    While it appears complex this is just the implementation for a variable in simply typed lambda calculus just hidden in the names. We move on to the implementation of type substitution
    
    \subsubsection{Type Substitution}
    
    Because of the dependent product and sum types we must define a substitution of expressions into types. In this no type can depend on a general expression and so we only have substitution defined under two situations either when the expression \verb|e| reduces to a name which will have no effect. The other case is when the variable \verb|x| is not free in the second type. In which case again the substitution will have no effect. Despite this we define a substitution function \verb|subst_ty| similar to the expression substitution with notation \verb|"[e/x]T"| for when \verb|e| is a name. The implementation is modelled on expression substitution.
    
    There is a second substitution of types necessary for the implementation of iso-recursive types. this is substitution for type variables similar to how basic lambda calculus substitutes into expression variables. This substitution is denoted by the notation \verb|"[X to U]T"| meaning substitute all occurences of \verb|X| for \verb|U| in type \verb|T|. This substitution is necessary for typing the ico-recursive types.
    
    \subsubsection{Expression Typing}
    
    To begin We use the notation \verb+"G|--S,e::T"+ to denote a term \verb|e| being of type \verb|T| under the type context \verb|G| and the tag store \verb|S|. We begin by discussing the rules that do not follow directly from the simply typed lambda calculus beginning with the \verb|t_app| rule for dependent function type as below and continuing the cases explored in the ToTO language and then progressing to other ones not typically in extensions of the simply typed lambda calculus.
    
    \begin{verbatim}
        | t_app : forall G S e1 (e2:name) x T T',
            wellformed_ty T ->
            wellformed_ty T' ->
            G |-- S , e1 :: Prod[ x : T ], T' ->
            G |-- S , e2 :: T ->
            G |-- S , <{ e1 e2 }> :: <{{[e2/x]T'}}>
    \end{verbatim}
    
    The only difference in this from the regular simply typed lambda calculus is the type substitution into the \verb|T'| type in the conclusion.  The subsumption rule implementation requires no explanation as it is a simple translation into Coq. Moving on we discuss all the rules relating to the tags and the types of tag creation and interaction terms.
    
    \begin{verbatim}
        | t_cvar : forall G S c T,
            wellformed_ty T ->
            S c = Some T ->
            G |-- S , c :: T
    \end{verbatim}
    
    The first rule implemented here \verb|t_cvar| is an implementation of \textit{CVar} looks up a tag value \verb|c| in the tag store and then generates the type judgement. The lookup in the context is generated as checking that the function \verb|S| maps \verb|c| to some type \verb|T|. The next implementations of interest are the implementations of \textit{CCLS-I}, \textit{Tag-I} and \textit{Extract}. In the ToTO language the depend on having the tag type $Tag(\tau)$ Which in the implementation of types is two separate types \verb|Tag[T]| and \verb|Tag[T]Extends(n)| Therefore we need the type judgements to be separated into two examples as in the case of \textit{Extract} it is implemented by \verb|t_ext1| and \verb|t_ext2| with one case handling each tag type as below.
    
    \begin{verbatim}
        | t_ext1 : forall G S e (n:name) T,
            wellformed_ty T ->
            G |-- S , e :: Tagged( n ) ->
            G |-- S , n :: Tag[ T ] ->
            G |-- S , Extract{ e } :: T
        | t_ext2 : forall G S e (n:name) n' T,
            wellformed_ty T ->
            G |-- S , e :: Tagged( n ) ->
            G |-- S , n :: Tag[ T ]Extends( n') ->
            G |-- S , Extract{ e } :: T
    \end{verbatim}
    
    We next examine the \textit{Match} rule implementation. The implementation is mostly self explanatory, The only new part is the hypothesis $\Gamma\vdash_\Sigma$ \verb|tagged| $n \updownarrow$ \verb|tagged| $n'$. Therefore we need to implement a proposition \verb|mutual_supertype| which is to determine if two types share a common supertype which is done below using the existential quantifier as below
    
    \begin{verbatim}
Definition mutual_supertype (G:context) (S:context) (T1:ty) (T2:ty) : Prop :=
    (exists T, (E ! G |- S , T1 <: T) /\ (E ! G |- S , T2 <: T)).
    \end{verbatim} 
    
    This is then used in the implementation of the \verb|t_match| rule.
    
    \begin{verbatim}
        | t_match : forall G S x (n:name) (n':name) e1 e2 e3 T,
            wellformed_ty T ->
            G |-- S , e1 :: Tagged( n' ) ->
            mutual_supertype G S <{{Tagged(n)}}> <{{Tagged(n')}}>->
            (c_update G x (Some <{{ Tagged( n ) }}>)) |-- S , e2 :: T ->
            G |-- S , e3 :: T ->
            G |-- S , Match{ e1 }( n )( x ){ e2 }{ e3 } :: T
    \end{verbatim}
    
    Other than this it is a direct implementation of the rule.
    
    The record term types are defined using three rules from \cite{Pierce:SF2} the \verb|t_rnils| judgement is very simple and just states that an empty record has type of an empty record. The \verb|t_rcons| is implemented below.
    
    \begin{verbatim}
    | t_rcons : forall G S i1 T1 e1 e2 er R,
        wellformed_ty T1 ->
        wellformed_ty R ->
        G |-- S, e1 :: T1 ->
        G |-- S, <{er}> :: <{{R}}> ->
        record_ty R ->
        record_tm er ->
        G |-- S, <{ i1 ;= e1 ;; er}> :: <{{ i1 ; T1 ;; R}}>
    \end{verbatim}
    
    The bulk of this is checking that types and terms are wellformed records. The Implementation says that a given record expression has the type of a record if the head and tail of the records have the appropriate types. The \verb|t_proj| judgements only new part is the use of the \verb|tyLookup| function which checks that it is possible to project a given variable from the record.
    
    The Sum type implementation is straightforward from the rules just remembering the substitution because the type is dependent.
    
    The \verb|t_fold| and \verb|t_unfold| judgements they are listed below.
    
    \begin{verbatim}
        | t_fld : forall G S (t:string) e T,
            wellformed_ty T ->
            G |-- S, e :: <{{mu(t):T}}> ->
            G |-- S, Fold[mu(t):T]{e} :: [t to mu(t):T]T
        | t_unfld : forall G S e (t:string) T,
            wellformed_ty T ->
            G |-- S, e :: <{{mu(t):T}}> ->
            G |-- S, <{Unfold{e} }>::<{{[t to mu(t):T]T}}>
    \end{verbatim}
    
    These make use of the second type of type substitution with substitution for type variables that was defined earlier. The typing rules are based on Pierce's Types and Programming languages book \cite{10.5555/509043} and the full implementation can be seen in \ref{app:type}.
    
    \subsubsection{Dynamics}
    
    In discussing, we must first return to the dynamic hierarchical store to remember all tags generated by the program; this storing prevents a function called an arbitrary number of times from generating an arbitrary number of tags. The dynamics of the language are in value judgements and small step reductions. The value judgements are represented with the notation.
    
    \begin{verbatim}
        "S ! e val"
    \end{verbatim} 
    
    which means that under the context of the store type \verb|S| that e is a value. The small step reductions are encoded as 
    
    \begin{verbatim}
        S ! e |->S'\e'
    \end{verbatim}
    
    This represents that under the context of the store \verb|S| the expression \verb|e| reduces to \verb|e'| while simultaneously extending the store to some \verb|S'|.
    
    We begin by examining the value judgements. The implementation is listed below in figure \ref{fig:vimp}.
    
    \begin{figure}[t]
        \begin{verbatim}
        Inductive value : store -> tm -> Prop :=
            | v_new : forall S e n,
                S ! e val ->
                S ! <{New{e}(n)}> val
            | v_lam : forall S x e T,
                S ! <{/x:T,e}> val
            | v_c : forall x S,
                x ins S ->
                S ! x val
            | v_rcd : forall f1 f2 e1 e2 R S,
                record_tm R ->
                S ! e1 val->
                S ! <{f2;=e2;;R}> val->
                S ! <{f1;=e1;;f2;=e2;;R}> val
            | v_prod : forall S e1 e2,
                S ! e1 val->
                S ! e2 val->
                S ! <{<e1,e2>}> val
            | v_unit : forall S,
                S ! <{< >}> val
            where "S '!' e 'val'" := (value S e).
        \end{verbatim}
        \caption{Value Judgements}
        \label{fig:vimp}
    \end{figure}
    
    There's not a lot to explain about the value judgement implementation the only notable part is the \verb|v_c| rule which checks if a given tag is in the hierarchical store.
    
    \begin{figure}
        \begin{verbatim}
  | r_cls : forall S x T,
      x nin S ->
      S ! <{NewTag[T]}> |-> <{{{x ---> Ep ;; S}}}> \ x
  | r_ccls : forall S x x' p T,
      x' nin S ->
      <{{{x--->p ;;S}}}> ! <{SubTag[T](x)}> |-> <{{{x'--->(x--->p);;(x---> p;;S)}}}> \ x'
  | r_new : forall S S' e e' n,
      S ! e|->S'\e' ->
      S !<{New{e}(n)}>|->S'\<{New{e}(n)}>
  | r_match : forall S S' e e' e2 e3 x y,
      S ! e|-> S' \ e' ->
      S ! <{Match{e}(x)(y){e2}{e3} }>|->S'\<{Match{e'}(x)(y){e2}{e3} }>
  | r_matchsuc : forall S x p e x' y e2 e3,
      <{{{x--->p;;S}}}>! e val ->
      x' inp <{{{x--->p}}}> ->
      <{{{x--->p;;S}}}>!<{Match{New{e}(x)}(x')(y){e2}{e3} }>|->
        <{{{x--->p;;S}}}>\<{[y:=New{e}(x)]e2}>
  | r_matchfail : forall S x p e x' y e2 e3,
      <{{{x--->p;;S}}}>! e val ->
      x' ninp <{{{x--->p}}}> ->
      <{{{x--->p;;S}}}>!<{Match{New{e}(x)}(x)(y){e2}{e3} }>|-><{{{x--->p;;S}}}>\e3
  | r_untag1 : forall S S' e e',
      S ! e |-> S' \ e' ->
      S ! <{Extract{e} }>|->S'\<{Extract{e'} }>
  | r_untag2 : forall S e x,
      S ! e val ->
      S ! <{Extract{New{e}(x)} }>|-> S\e
	    \end{verbatim}
        \caption{New small step reduction implementation}
        \label{fig:ssimp}
    \end{figure}
    
    From that we move along to the small step evaluation rules. We include the implementation of all the new rules here in figure \ref{fig:ssimp}. In \verb|r_cls| and \verb|r_ccls| use the previously defined predicates for determining the existence of a tag value in the store to then update the store and evaluate to the tag value. In particular the \verb|r_ccls| updates the store with a given tag as a supertag.
    
    The next important implementations are \verb|r_matchsuc| and \verb|r_matchfail| which implement the dynamics around dynamically matching tags. In these implementations the only thing to note is the use of the path inclusion checking to determine if \verb|x'| is a supertag of \verb|x|. The remainder of the evaluation steps for the tagging expressions are fairly straightforward direct implementations of the dynamics in the core language.
    
    For the remainder of the evaluation rules most are direct implementations of the evaluation rules presented in Pierce \cite{10.5555/509043} and so are not expanded on to view the entire predicate function see \ref{app:red}. The only exceptions are the record evaluation rules because of the recursive implementation of records. The projection rule called by Pierce \textit{E-ProjRcd} is implemented using the rule \verb|r_projrcd| and utilises the \verb|tmlookup| function to check if a given variable is in the record. similar to how \verb|tylookup| is used in \verb|t_proj|. The other two rules \verb|r_rcdhead| and \verb|r_rcdtail| implement evaluation of the head and tail of a record.
    
    \subsection{Type Soundness}
    
    Before we can discuss the statements of the type soundness theorems there are two more relations to implement the first is the subcontext relation $\Sigma'\subset \Sigma$. If we treat a context almost as a set we just need that for all $c$ and $\tau$ that $c \sim \tau\in \Sigma'$ implies $c \sim \tau \in \Sigma$. We implement this with the below predicate.
    
    \begin{verbatim}
        Definition subcontext (S:context) (S':context) : Prop :=
            forall c T, S c = Some T -> S' c = Some T.
    \end{verbatim}
    
    The second relation is the relation checking that the static tag context $\Sigma$ and the dynamic hierarchical tag store \verb|S| have the same associated tags and subtags. This is implemented from the supplementary material of the theory of tagged objects as:
    
    \begin{verbatim}
    Inductive storecontext : context -> store -> Prop :=
        | sc_empty :
            storecontext E store_empty
        | sc_step : forall F S x T,
            storecontext (c_update F x (Some T)) (store_cons <{{{x--->Ep}}}> S).
    \end{verbatim}
    
    We now have all the necessary parts to state the type soundness theorems.
    
    \subsubsection{Preservation}
    
    The Preservation theorem is implemented as below
    \begin{verbatim}
        Theorem Preservation : forall G F S S' e e' T,
            G |-- F , e :: T ->
            S ! e |-> S' \ e'->
            (exists F',(subcontext F' F)/\(G |-- F' , e' :: T)).
    \end{verbatim}
    
    \subsubsection{Progress}
    
    The Progress theorem is implemented as below
    \begin{verbatim}
        Theorem Progress : forall F e e' T S S',
            E |-- F, e :: T ->
            storecontext F S ->
            (S ! e val) \/ (S!e|-> S'\e').
    \end{verbatim}
    
    \section{RELATED WORK}
        
    The core language presented in this report began as a core type theory modelling class based object oriented languages. Many of the largest programming languages such as Java, C++ the object oriented programming is class based. In many class based object oriented language models there is a static class hierarchy which is not flexible. The source language of \cite{lee_et_al:LIPIcs.ECOOP.2015.174} being modelled supports a dynamic class hierarchy other research into dynamic typing and the interaction with static typing is in \cite{10.1145/2398857.2384674}. Other research into dynamically typed object oriented languages is in \cite{DBLP:journals/corr/EngelmannO15}, Which describes a complete Hoare logic on a simple dynamically typed object oriented languages. 
    
    \section{DISCUSSION}
    
    There is not any major difference from the implementations.
    
    Implementation of the syntax and semantics to a mechanised theorem prover such as Coq would allow mechanised reasoning about the type system. In particular mechanised proofs of the type soundness theorems allows confidence that the type system is valid. However this does not mean that the encoded is consistent with the type system on paper.
    
    For example this encoding has differences to the original ToTO core language in that it implements a fix expression as opposed to a recursive let binding and so even if the type soundness theorems were proved this would not ensure the original core language was correct.
    
    Because we have neither encoded nor proved any theorems about the type system encoded I can not claim that the type system is correct or fully consistent with the ToTO type system. In particular there is not an entirely dependent product or sum type since the substitution of expressions into types is limited to substituting for names in the implementation.
    
    The issues of type checking decidability or interactions with multiple representations have not beed addressed and are left as future work~\cite{mackay:2020:popl,10.1145/2600176.2600194}. One may also want to examine the interaction with memory management approaches dealing with aliasing~\cite{potanin:2002:checking-ownership-and-confinement,potanin:2013:icse,potanin:2004:ftfjp,craig:2018:effects,potanin:2005:scopes}.
    
    The use of Coq proof assistant for the implementation allowed the use of Coq's predicate logic system which means we can bypass the Coq type checker's insistence that a recursive function must terminate. This along with not having the necessity to define a case for every possible input means we can effectively encode rules such as the value judgement rules which does not make a judgement on every possible expression.
    
    \section{CONCLUSION AND FUTURE WORK}
    
    We have attempted present an encoding of the type theory of the Core language of the theory of tagged objects into Coq Proof Assistant. We have also encoded declarations of the type soundness theorems Preservation and Progress.
    
    In future it would be beneficial to mechanise the proofs of type soundness because then we can be confident that the type system encoded is a valid type system. The original ToTO paper includes a source class based object oriented language implementing this language and the language translations would allow more direct reasoning about class based languages. Lastly extending the tag system to allow multiple inheritance.
    
    \bibliographystyle{plain}

    \pagebreak
    
    \appendix
    \renewcommand{\thesection}{Appendix \Alph{section}.}
    \section{Subtyping Implementation}
    \label{app:sub}
    \begin{verbatim}
        Inductive subtype : context -> context -> context -> ty -> ty -> Prop :=
            | s_refl : forall D G S T, 
                wellformed_ty T ->
                D ! G |- S , T <: T
            | s_ : forall D G S T1 T2 T3,
                wellformed_ty T1 ->
                wellformed_ty T2 ->
                wellformed_ty T3 ->
                D ! G |- S , T1 <: T2 ->
                D ! G |- S , T2 <: T3 ->
                D ! G |- S , T1 <: T3
            | s_amb1 : forall D G S t t',
                wellformed_ty t ->
                wellformed_ty t' ->
                D t = Some t' ->
                D ! G |- S , t <: t'
            | s_amb2 : forall D G S (t:string) (t':string) T T',
                wellformed_ty t ->
                wellformed_ty t' ->
                wellformed_ty T ->
                wellformed_ty T' ->
                (c_update D t (Some (ty_unit t'))) ! G |- S , T <: T' ->
                D ! G |- S , <{{ mu( t ): T }}> <: <{{ mu( t' ): T' }}>
            | s_rcd1 : forall D G S i T R,
                wellformed_ty <{{ i ; T ;; R }}> ->
                D ! G |- S , <{{ i ; T ;; R }}> <: <{{ nil }}>
            | s_rcd2 : forall D G S i1 i2 T1 T2 R1 R2,
                wellformed_ty <{{ i1 ; T1 ;; R1 }}> ->
                wellformed_ty <{{ i2 ; T2 ;; R2 }}> ->
                D ! G |- S , T1 <: T2 ->
                D ! G |- S , R1 <: R2 ->
                D ! G |- S , <{{ i1 ; T1 ;; R2 }}> <: <{{ i2 ; T2 ;; R2 }}>
            | s_rcd3 : forall D G S i1 i2 T1 T2 R,
                wellformed_ty <{{ i1 ; T1 ;; i2 ; T2 ;; R }}> ->
                i1 <> i2 ->
                D ! G |- S , <{{ i1 ; T1 ;; i2 ; T2 ;; R }}> <: <{{ i2 ; T2 ;; i1 ; T1 ;; R }}>
            | s_app : forall D G S x T1 T2 T3 T4,
                wellformed_ty T1 ->
                wellformed_ty T2 ->
                wellformed_ty T3 ->
                wellformed_ty T4 ->
                D ! G |- S , T3 <: T1 ->
                D ! (c_update G x (Some T3)) |- S , T2 <: T4 ->
                D ! G |- S , Prod[ x : T3 ], T2 <: Prod[ x : T3 ], T4
            | s_tag2 : forall D G S n n' T,
                wellformed_ty T ->
                D ! G |- S , Tagged( n ) <: Tagged( n' ) ->
                D ! G |- S , Tag[ T ]Extends( n ) <: Tag[ T ]Extends( n' )
            | s_tag3 : forall D G S n T,
                wellformed_ty T ->
                D ! G |- S , Tag[ T ]Extends( n ) <: Tag[ T ]
        where "D '!' G '|-' S ',' t '<:' t'" := (subtype D G S t t').
    \end{verbatim}
    \section{Expression Typing Implementation}
    \label{app:type}
    \begin{verbatim}
        Inductive type : context -> context -> tm -> ty -> Prop :=
            | t_app : forall G S e1 (e2:name) x T T',
                wellformed_ty T ->
                wellformed_ty T' ->
                G |-- S , e1 :: Prod[ x : T ], T' ->
                G |-- S , e2 :: T ->
                G |-- S , <{ e1 e2 }> :: <{{[e2/x]T'}}>
            | t_sub :  forall G S e T T',
                wellformed_ty T ->
                wellformed_ty T' ->
                G |-- S , e :: T ->
                E ! G |- S , T <: T' ->
                G |-- S , e :: T'
            | t_cvar : forall G S c T,
                wellformed_ty T ->
                S c = Some T ->
                G |-- S , c :: T
            | t_clsI : forall G S T,
                wellformed_ty T ->
                G |-- S , NewTag[ T ] :: Tag[ T ]
            | t_cclsI1 : forall G S (n:name) T T',
                wellformed_ty T ->
                wellformed_ty T' ->
                G |-- S , n :: Tag[ T ] ->
                E ! G |- S , T' <: T ->
                G |-- S , SubTag[ T' ]( n ) :: Tag[ T' ]Extends( n )
            | t_cclsI2 : forall G S (n:name) n' T T',
                wellformed_ty T ->
                wellformed_ty T' ->
                G |-- S , n :: Tag[ T ]Extends(n') ->
                E ! G |- S , T' <: T ->
                G |-- S , SubTag[ T' ]( n ) :: Tag[ T' ]Extends( n )
            | t_tag1 : forall G S (n:name) e T,
                wellformed_ty T ->
                G |-- S , n :: Tag[ T ] ->
                G |-- S , e :: T ->
                G |-- S , New{ e }(n) :: Tagged( n )
            | t_tag2 : forall G S (n:name) n' e T,
                wellformed_ty T ->
                G |-- S , n :: Tag[ T ]Extends(n') ->
                G |-- S , e :: T ->
                G |-- S , New{ e }(n) :: Tagged( n )
            | t_match : forall G S x (n:name) (n':name) e1 e2 e3 T,
                wellformed_ty T ->
                G |-- S , e1 :: Tagged( n' ) ->
                mutual_supertype G S <{{Tagged(n)}}> <{{Tagged(n')}}>->
                (c_update G x (Some <{{ Tagged( n ) }}>)) |-- S , e2 :: T ->
                G |-- S , e3 :: T ->
                G |-- S , Match{ e1 }( n )( x ){ e2 }{ e3 } :: T
            | t_ext1 : forall G S e (n:name) T,
                wellformed_ty T ->
                G |-- S , e :: Tagged( n ) ->
                G |-- S , n :: Tag[ T ] ->
                G |-- S , Extract{ e } :: T
            | t_ext2 : forall G S e (n:name) n' T,
                wellformed_ty T ->
                G |-- S , e :: Tagged( n ) ->
                G |-- S , n :: Tag[ T ]Extends( n') ->
                G |-- S , Extract{ e } :: T
            (* Sum Type *)
            | t_sum_1 : forall G S (e1:name) e2 T1 T2 x,
                wellformed_ty T1 ->
                wellformed_ty T2 ->
                G |-- S , e1 :: T1 ->
                G |-- S , e2 :: [e1/x]T2 ->
                G |-- S , <e1,e2> :: Sum[x:T1]T2
            | t_e1 : forall G S e x T1 T2,
                wellformed_ty T1 ->
                wellformed_ty T2 ->
                G |-- S , e :: Sum[x:T1]T2 ->
                G |-- S , Fst{e} :: T1
            | t_e2 : forall G S (e:name) x T1 T2,
                wellformed_ty T1 ->
                wellformed_ty T2 ->
                G |-- S , e :: Sum[x:T1]T2 ->
                G |-- S, Snd{e} :: [Fst(e)/x]T2
            (* Record Typing *)
            | t_rcons : forall G S i1 T1 e1 er R,
                wellformed_ty T1 ->
                wellformed_ty R ->
                G |-- S, e1 :: T1 ->
                G |-- S, <{er}> :: <{{R}}> ->
                record_ty R ->
                record_tm er ->
                G |-- S, <{ i1 ;= e1 ;; er}> :: <{{ i1 ; T1 ;; R}}>
            | t_proj : forall G S i (f:string) T R,
                wellformed_ty T ->
                wellformed_ty R ->
                record_ty R ->
                G |-- S, f :: R ->
                tylookup i R = Some T ->
                G |-- S, <{i proj f}> :: T
            | t_rnil : forall G S,
                G |-- S, <{nil}> :: <{{nil}}>
            (* Let Bindings *)
            | t_let : forall G S e1 e2 T1 T2 x,
                wellformed_ty T1 ->
                wellformed_ty T2 ->
                G |-- S, e1 :: T1 ->
                (c_update G x (Some T1)) |-- S, e2 :: T2 ->
                G |-- S, <{Let x be e1 in e2}> :: T2
            | t_fix : forall G S e T x,
                wellformed_ty T ->
                G |-- S, e :: <{{Prod[x:T],T}}> ->
                G |-- S, <{Fix{e} }> :: T
            (* Recursive *)
            | t_fld : forall G S (t:string) e T,
                wellformed_ty T ->
                G |-- S, e :: <{{mu(t):T}}> ->
                G |-- S, Fold[mu(t):T]{e} :: [t to mu(t):T]T
            | t_unfld : forall G S e (t:string) T,
                wellformed_ty T ->
                G |-- S, e :: <{{mu(t):T}}> ->
                G |-- S, <{Unfold{e} }>::<{{[t to mu(t):T]T}}>
        where "G '|--' S ',' e '::' T" := (type G S e T).
    \end{verbatim}
    \section{Reduction Implementation}
    \label{app:red}
        \begin{verbatim}
    Inductive reduction : store -> store -> tm -> tm -> Prop :=
        | r_cls : forall S x T,
            x nin S ->
            S ! <{NewTag[T]}> |-> <{{{x ---> Ep ;; S}}}> \ x
        | r_ccls : forall S x x' p T,
            x' nin S ->
            <{{{x--->p ;;S}}}> ! <{SubTag[T](x)}> |->
              <{{{x'--->(x--->p);;(x---> p;;S)}}}> \ x'
        | r_new : forall S S' e e' n,
            S ! e|->S'\e' ->
            S !<{New{e}(n)}>|->S'\<{New{e}(n)}>
        | r_match : forall S S' e e' e2 e3 x y,
            S ! e|-> S' \ e' ->
            S ! <{Match{e}(x)(y){e2}{e3} }>|->S'\<{Match{e'}(x)(y){e2}{e3} }>
        | r_matchsuc : forall S x p e x' y e2 e3,
            <{{{x--->p;;S}}}>! e val ->
            x' inp <{{{x--->p}}}> ->
            <{{{x--->p;;S}}}>!<{Match{New{e}(x)}(x)(y){e2}{e3} }>|->
              <{{{x--->p;;S}}}>\<{[y:=New{e}(x)]e2}>
        | r_matchfail : forall S x p e x' y e2 e3,
            <{{{x--->p;;S}}}>! e val ->
            x' ninp <{{{x--->p}}}> ->
            <{{{x--->p;;S}}}>!<{Match{New{e}(x)}(x)(y){e2}{e3} }>|->
              <{{{x--->p;;S}}}>\e3
        | r_untag1 : forall S S' e e',
            S ! e |-> S' \ e' ->
            S ! <{Extract{e} }>|->S'\<{Extract{e'} }>
        | r_untag2 : forall S e x,
            S ! e val ->
            S ! <{Extract{New{e}(x)} }>|-> S\e
        (* Records *)
        | r_projrcd : forall S i er vi,
            S ! er val ->
            tmlookup i er = Some vi ->
            S ! <{er proj i}> |->S\vi
        | r_rcdhead : forall S i e1 e1' er,
            S ! e1|->S\e1' ->
            S ! <{i;=e1;;er}>|->S\<{i;=e1';;er}>
        | r_rcdtail : forall S i e1 er er',
            S ! e1 val ->
            S ! er |-> S\er' ->
            S ! <{i;=e1;;er}>|->S\<{i;=e1;;er'}>
        (* let bindings *)
        | r_letv : forall S x v e,
            S ! v val ->
            S ! <{Let x be v in e}> |-> S \ [x:=v]e
        | r_let : forall S x e1 e1' e2,
            S ! e1 |-> S \ e1' ->
            S ! <{Let x be e1 in e2}> |-> S \ <{Let x be e1' in e2}>
        | r_fixb : forall S x T e,
            S ! Fix{/x:T,e} |-> S \ [x := Fix{/x:T,e}]e
        | r_fix : forall S e e',
            S ! e |-> S \ e' ->
            S ! Fix{e} |-> S \ Fix{e'}
        (* Recursive *)
        | r_unfldfld : forall S T v,
            S ! v val ->
            S ! <{Unfold{Fold[T]{v}} }> |-> S \ v
        | r_fld : forall S T e1 e1',
            S ! e1 |-> S \ e1' ->
            S ! Fold[T]{e1} |-> S \ Fold[T]{e1'}
        | r_unfld : forall S e1 e1',
            S ! e1 |-> S \ e1' ->
            S ! Unfold{e1} |-> S \ Unfold{e1'}
        (* Sums *)
        | r_pairv1 : forall S v1 v2,
            S ! <v1,v2> val ->
            S ! Fst{<v1,v2>} |-> S \ v1
        | r_pairv2 : forall S v1 v2,
            S ! <v1,v2> val ->
            S ! Snd{<v1,v2>} |-> S \ v2
        | r_proj1 : forall S e e',
            S ! e |-> S \ e' ->
            S ! Fst{e} |-> S \ Fst{e'}
        | r_proj2 : forall S e e',
            S ! e |-> S \ e' ->
            S ! Snd{e} |-> S \ Snd{e'}
        | r_pair1 : forall S e1 e1' e2,
            S ! e1 |-> S \ e1' ->
            S ! <e1,e2> |-> S \ <e1',e2>
        | r_pair2 : forall S e1 e2 e2',
            S ! e2 |-> S \ e2' ->
            S ! <e1,e2> |-> S \ <e1,e2'>
        (* Products *)
        | r_appabs : forall S x T v e,
            S ! v val ->
            S ! <{(/x:T,e) v}> |-> S \ [x:=v]e
        | r_app1 :forall S e1 e1' e2,
            S ! e1 |-> S \ e1' ->
            S ! <{e1 e2}> |-> S \ <{e1' e2}>
        | r_app2 :forall S v e2 e2',
            S ! v val ->
            S ! e2 |-> S \ e2' ->
            S ! <{v e2}> |-> S \ <{v e2'}>
    where "S '!' e '|->' S' '\' e'" := (reduction S S' e e').
        \end{verbatim}
\end{document}